\documentstyle[preprint,prl,aps,epsf]{revtex}
\begin{document}

\tightenlines

\preprint{\vbox{\hbox{BNL-64631}
\hbox{KEK Preprint 97-115}
\hbox{PRINCETON/HEP/97-12}
\hbox{TRI--PP--97--28}}}
\title{Evidence for
the Decay $K^+ \to \pi^+ \nu \bar\nu$}


\author{S.~Adler$^1$, M.S.~Atiya$^1$, I-H.~Chiang$^1$, M.V.~Diwan$^1$,
  J.S.~Frank$^1$, J.S.~Haggerty$^1$,\\ S.H.~Kettell$^1$, 
  T.F.~Kycia$^1$, K.K.~Li$^1$, L.S.~Littenberg$^1$, C.~Ng$^1$\cite{chi},
  A.~Sambamurti$^1$\cite{AKS}, \\ A.~Stevens$^1$,
  R.C.~Strand$^1$,C.~Witzig$^1$, T.K.~Komatsubara$^2$, M.~Kuriki$^2$,
  N.~Muramatsu$^2$, S.~Sugimoto$^2$, T.~Inagaki$^3$, S.~Kabe$^3$,
  M.~Kobayashi$^3$, Y.~ Kuno$^3$, T.~Sato$^3$, T.~Shinkawa$^3$,\\
  Y.~Yoshimura$^3$, Y.~Kishi$^4$, T.~Nakano$^4$,
  M.~Ardebili$^5$, A.O.~Bazarko$^5$, M.R.~Convery$^5$,
  M.M.~Ito$^5$, D.R.~Marlow$^5$,
  R.A.~McPherson$^5$, P.D.~Meyers$^5$,\\ F.C.~Shoemaker$^5$,
  A.J.S.~Smith$^5$, J.R.~Stone$^5$, M.~Aoki$^6$,
  E.W.~Blackmore$^6$,\\ P.C.~Bergbusch$^6$, D.A.~Bryman$^6$, 
A.~Konaka$^6$,
  J.A.~Macdonald$^6$, J.~Mildenberger$^6$, \\T.~Numao$^6$,
  P.~Padley$^6$, J.-M.~Poutissou$^6$,
R.~Poutissou$^6$,
  G.~Redlinger$^6$, \\J.~Roy$^6$,
  A.S.~Turcot$^6$,
P.~Kitching$^7$ and
   R.~Soluk$^7$\\ (E787 Collaboration) }

\address{ $^1$Brookhaven National Laboratory, Upton, New York 11973\\
  $^2$High Energy Accelerator Research Organization (KEK),
  Tanashi-branch, Midoricho, Tanashi, Tokyo 188, Japan \\ $^3$High
  Energy Accelerator Research Organization (KEK), Oho, Tsukuba,
  Ibaraki 305, Japan \\ $^4$Department of Physics, Osaka University,
  Machikaneyama, Toyonaka, Osaka 560, Japan \\ $^5$Joseph Henry
  Laboratories, Princeton University, Princeton, New Jersey 08544 \\
  $^6$ TRIUMF, 4004 Wesbrook Mall, Vancouver, British Columbia,
  Canada, V6T 2A3\\
$^7$ Centre for Subatomic Research, University of Alberta, Edmonton,
Alberta, T6G 2N5}

\date{August 29, 1997}
\maketitle

\begin{abstract}
An event consistent with the signature expected for the rare kaon
decay $K^+ \to \pi^+ \nu \bar\nu$ ~ has been observed.  In the pion momentum region examined,
$211<P<230$ MeV/c, the backgrounds are estimated to contribute
$0.08\pm0.03$ events. If the event is due to $K^+ \to \pi^+ \nu \bar\nu$, the branching
ratio is $4.2^{+9.7}_{-3.5} \times 10^{-10}$.
\end{abstract}



\input psfig


The decay $K^+ \to \pi^+ \nu \bar\nu$~ has attracted interest due to
its sensitivity to $|V_{td}|$, the coupling of top to down quarks in
the Cabibbo-Kobayashi-Maskawa quark mixing matrix.  Theoretical
uncertainty in the branching ratio is minimal because the decay rate
depends on short distance physics and because the hadronic matrix
element can be extracted from the well-measured decay $K^+\to\pi^0
e^+\nu$.  After next-to-leading-logarithmic analysis of QCD
effects\cite{BB3}, calculation of isospin breaking, phase space
differences and other small corrections to the hadronic matrix
element~\cite{2mar}, and calculation of two-electroweak-loop
effects~\cite{twoloop}, the intrinsic uncertainty is only about
7\%\cite{bf}.  Based on current knowledge of Standard Model (SM)
parameters, the branching ratio $B(K^+\to\pi^+\nu\bar\nu)$ is expected
to be in the range $0.6 - 1.5 \times 10^{-10}$\cite{wdbll}.
Long-distance contributions to the branching ratio ({\it{i.e.}}
meson, photon exchange) appear to be negligible
($10^{-13}$)\cite{rhl,longd}.  Since $K^+ \to \pi^+ \nu \bar\nu$~ is a
flavor changing neutral current process that is highly suppressed in
the SM, it also serves as a hunting ground for non-SM physics.  The
signature $K^+ \to \pi^+ $ `nothing'\cite{rhl,BSM,familon} includes
$K^+ \to \pi^+ \nu \bar \nu$ with non-SM intermediate states (such as
virtual supersymmetric particles), $K^+ \to \pi^+ \nu \bar \nu'$ (a
lepton flavor violating final state), $K^+ \to \pi^+ X^0 X^{0'}$ where
$X^0$ and $X^{0'}$ are not neutrinos, and $K^+ \to \pi^+ X^0$ where
$X^0$ is a single, non-interacting particle. Initial results from the
E787 experiment\cite{e787nim} at the Alternating Gradient Synchrotron
(AGS) of Brookhaven National Laboratory gave 90\% confidence level
(CL) upper limits $B$($K^+ \to \pi^+ \nu \bar\nu$) $< 2.4 \times
10^{-9}$ and $B$($K^+ \to \pi^+ X^0$)$< 5.2 \times 10^{-10}$ for a
massless $X^0$\cite{pnn96}. In this letter, we report on the analysis
of a new data sample with 2.4 times greater sensitivity, taken in 1995
using an upgraded beam and detector.

The signature for $K^+ \! \rightarrow \! \pi^+ \nu \overline{\nu}$ is
a $K^+$ decay to a $\pi^+$ of momentum $P<227$ MeV/$c$ and no other
observable product. Definitive observation of this signal requires
suppression of all backgrounds to well below the sensitivity for the
signal and reliable estimates of the residual background levels.
Major background sources include the copious two-body decays $K^+ \!
\rightarrow \! \mu^+ \nu_\mu$ ($K_{\mu 2}$) with a 64\% branching
ratio and $P=236$ MeV/$c$ and $K^+ \!  \rightarrow \!  \pi^+ \pi^0$
($K_{\pi 2}$) with a 21\% branching ratio and $P=205$ MeV/$c$. The
only other important background sources are scattering of pions in the
beam and $K^+$ charge exchange (CEX) reactions resulting in decays
$K_L^0\to\pi^+ l^- \overline\nu$, where $l=e$ or $\mu$.  To suppress
the backgrounds, techniques were employed that incorporated redundant
kinematic and particle identification measurements and efficient
elimination of events with additional particles.

Kaons of 790~MeV/$c$~ were delivered to the experiment at a rate of
$7\times10^6$ per 1.6-s spill of the AGS. The kaon beam line (LESB3)
incorporated two stages of particle separation resulting in a pion
contamination of about 25\%.  The kaons were detected and identified
by \v{C}erenkov, tracking, and energy loss ($dE/dx$) counters.  About
20\% of the kaons passed through a degrader to reach a stopping 
target of 5-mm-square plastic scintillating fibers read out by
500-MHz CCD transient digitizers\cite{ccd}.  Measurements of the
momentum ($P$), range ($R$, in equivalent cm of scintillator) and
kinetic energy ($E$) of charged decay products were made using the
target, a central drift chamber\cite{utc}, and a cylindrical range
stack with 21 layers of plastic scintillator and two layers of straw
tube tracking chambers. Pions were distinguished from muons by
kinematics and by observing the $\pi \!  \rightarrow \! \mu \!
\rightarrow \!  e$ decay sequence in the range stack using 500-MHz
flash-ADC transient digitizers (TD)\cite{tds}.  Photons were detected
in a $4\pi$-sr calorimeter consisting of a
14-radiation-length-thick barrel detector made of lead/scintillator
and 13.5 radiation lengths of undoped CsI crystal detectors (also read
out using CCD digitizers) covering each end\cite{csi}. In addition,
photon detectors were installed in the extreme forward and backward
regions, including a Pb-glass \v{C}erenkov detector just upstream of
the target.  A 1-T solenoidal magnetic field was imposed on the
detector for the momentum measurements.

In the search for $K^+ \to \pi^+ \nu \bar\nu$, we required an
identified $K^+$ to stop in the target followed, after a delay of at
least 2 ns, by a single charged-particle track that was unaccompanied
by any other decay product or beam particle. This particle must
have been identified as a $\pi^+$ with $P$, $R$ and $E$ between the
$K_{\pi 2}$ and $K_{\mu 2}$ peaks.  A multilevel trigger selected
events with these characteristics for recording, and off-line analysis
further refined the suppression of backgrounds. To elude rejection,
$K_{\mu 2}$ and $K_{\pi 2}$ events would have to have been
reconstructed incorrectly in $P$, $R$ and $E$. In addition, any event
with a muon would have to have had its track misidentified as a pion
--- the most effective weapon here was the measurement of the $\pi \!
\rightarrow \!  \mu \!  \rightarrow \!  e$ decay sequence which
provided a suppression factor $10^{-5}$.  Events with photons, such as
$K_{\pi 2}$ decays, were efficiently eliminated by exploiting the full
calorimeter coverage. The inefficiency for detecting events with
$\pi^0$s was $10^{-6}$ for a photon energy threshold of about 1 MeV. A
scattered beam pion could have survived the analysis only by
misidentification as a $K^+$ and if the track were mismeasured as
delayed, or if the track were missed entirely by the beam counters
after a valid $K^+$ stopped in the target.  CEX background events
could have survived only if the $K_L^0$ were produced at low enough
energy to remain in the target for at least 2 ns, if there were no
visible gap between the beam track and the observed $\pi^+$ track, and
if the additional charged lepton went unobserved.

The data were analyzed with the goal of reducing the total expected
background to significantly less than one event in the final sample.
In developing the required rejection criteria (cuts), we took
advantage of redundant independent constraints available on each
source of background to establish two independent sets of cuts.  One
set of cuts was relaxed or inverted to enhance the background (by up
to three orders of magnitude) so that the other group could be
evaluated to determine its power for rejection.  For example, $K_{\mu
2}$ (including $K^+ \! \rightarrow \!  \mu^+ \nu_\mu \gamma$) was
studied by separately measuring the rejections of the TD particle
identification and kinematic cuts.  The background from $K_{\pi 2}$
was evaluated by separately measuring the rejections of the photon
detection system and kinematic cuts.  The background from beam pion
scattering was evaluated by separately measuring the rejections of the
beam counter and timing cuts.  Measurements of $K^+$ charge exchange
in the target were performed, which, used as input to Monte Carlo
studies, allowed the background to be determined.  Small correlations
in the separate groups of cuts were investigated for each background
source and corrected for if they existed.

The background levels anticipated with the final analysis cuts were
$b_{{K_{\mu 2}}} = 0.02 \pm 0.02$, $b_{{K_{\pi 2}}} = 0.03 \pm 0.02$,
$b_{Beam} = 0.02\pm 0.01$ and $b_{CEX} = 0.01 \pm 0.01$.  In total,
$b=0.08\pm 0.03$ background events were expected in the signal
region\cite{bkgold}.  Further confidence in the background estimates
and in the measurements of the background distributions near the
signal region was provided by extending the method described above to
estimate the number of events expected to appear when the cuts were
relaxed in predetermined ways so as to allow orders of magnitude
higher levels of all background types.  Confronting these estimates
with measurements from the full $K^+ \to \pi^+ \nu \bar\nu$~ data,
where the two sets of cuts for each background type were relaxed
simultaneously, tested the independence of the two sets of cuts.  At
approximately the $20 \times b$ level we observed 2 events where $1.6
\pm 0.6$ were expected, and at the level $150 \times b$ we found 15
events where $12 \pm 5$ were expected.  Under detailed examination,
the events admitted
by the relaxed cuts
were consistent with
being due to the known background sources.  Within the final signal
region, we still had additional background rejection capability.
Therefore, prior to looking in the signal region, we established
several sets of ever-tighter criteria which were designed to be used
only to interpret any events that fell into the signal region.

\begin{figure}
\begin{minipage}{0.24\linewidth}
\flushright{\psfig{figure=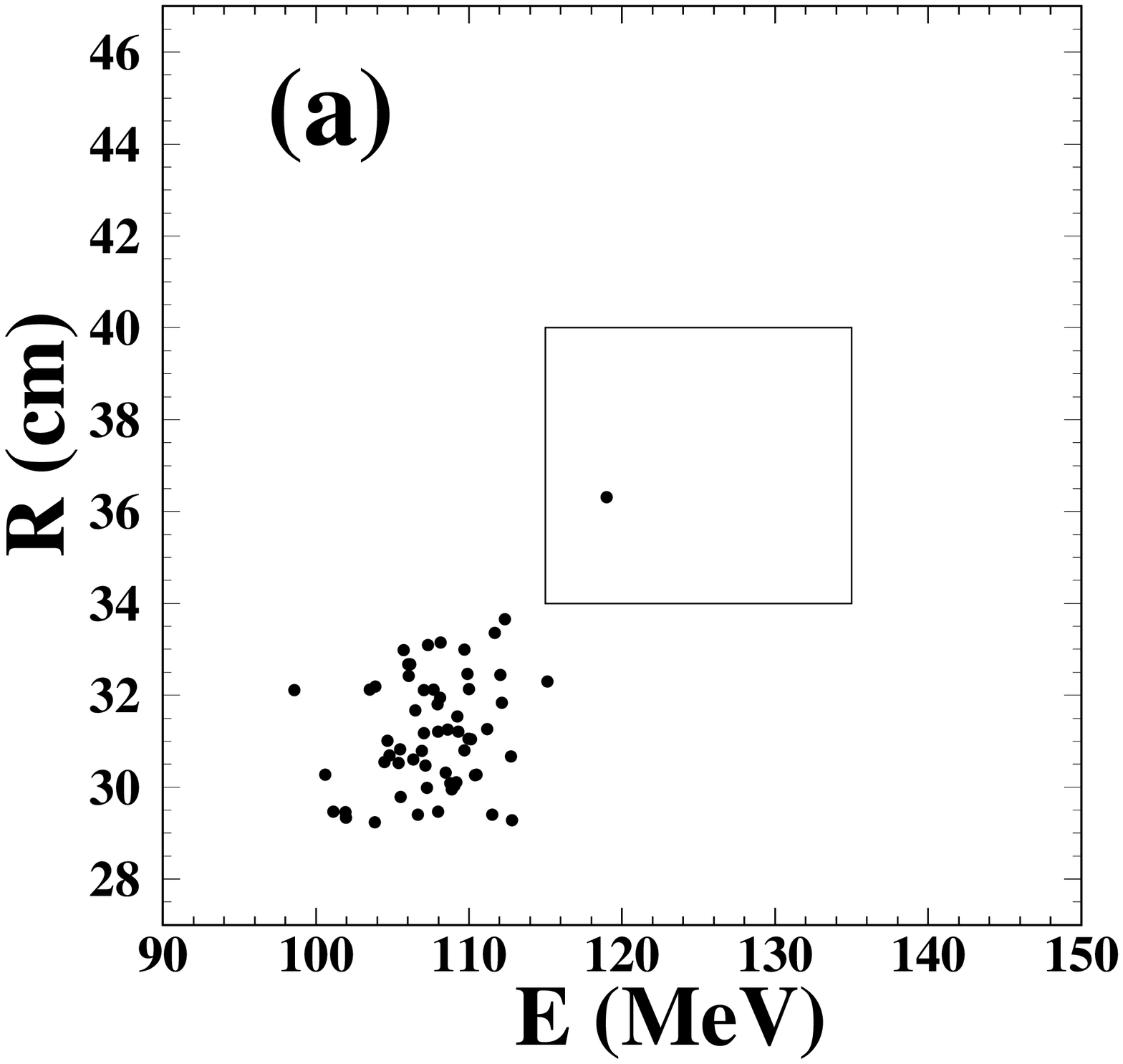,height=2.62in,width=2.62in}}
\end{minipage}\hfill
\begin{minipage}{0.24\linewidth}
\flushleft{\psfig{figure=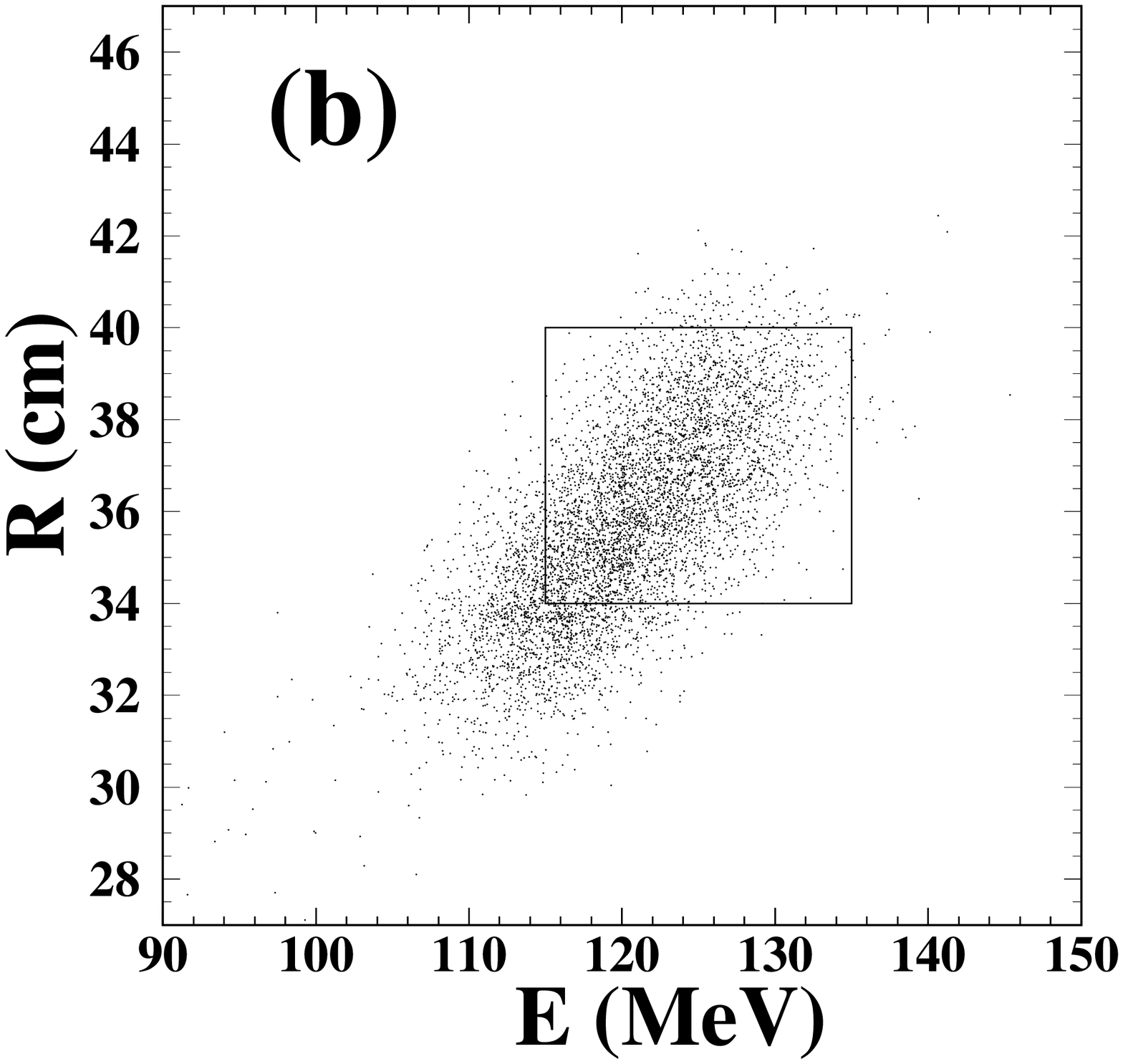,height=2.62in,width=2.62in}}
\end{minipage}\hfill
\caption{\label{data2}}{(a) Range ($R$) vs. energy ($E$) distribution
for the $K^+ \to \pi^+ \nu \bar\nu$~ data set with the final cuts
applied. The box enclosing the signal region contains a single
candidate event. (b) The Monte Carlo simulation of $K^+ \to \pi^+ \nu
\bar\nu$~ with the same cuts applied.}
\end{figure}

Figure~\ref{data2}(a) shows $R$ vs.~$E$ for the events surviving all
other analysis cuts. Only events with measured momentum in the
accepted region $211 \le P \le 230$ MeV/$c$ are plotted. The
rectangular box indicates the signal region specified as range $34 \le
R \le 40$ cm of scintillator (corresponding to $214 \le P_{\pi} \le
231$ MeV/$c$) and energy $115 \le E \le 135$ MeV ($213 \le P_{\pi} \le
236$ MeV/$c$) which encloses the upper 16.2\% of the $K^+ \to \pi^+
\nu \bar\nu$~ phase space.  One event was observed in the signal
region. The residual events below the signal region clustered at $E=
108$ MeV were due to $K_{\pi 2}$\ decays where both photons had been
missed. The number of these events is consistent with estimates of the
photon detection inefficiency.

\begin{figure}
\leavevmode 
\flushleft{\psfig{figure=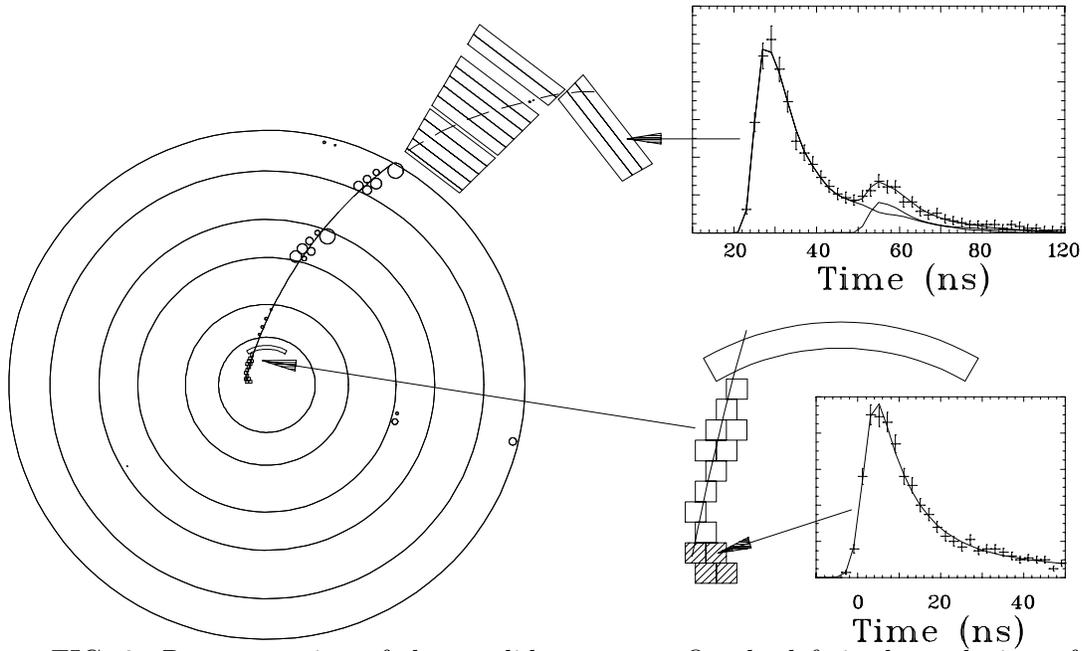,width=5.6in}}
\caption{\label{event1}Reconstruction of the candidate event. On the
left is the end view of the detector showing the track in the target,
drift chamber (indicated by drift-time circles), and range stack
(indicated by the layers that were hit). At the lower right is a
blowup of the target region where the hatched boxes are kaon hits, the
open boxes are pion hits, and the inner trigger counter hit is also
shown.  The pulse data sampled every 2 ns (crosses), in one of the
target fibers hit by the stopped kaon is displayed along with a fit
(curve) to the expected pulse shape. At the upper right of the figure
is the $\pi \to \mu$ decay signal in the range stack scintillator
layer where the pion stopped.  The crosses are the pulse data sampled
every 2 ns, and the curves are fits for the first, second and combined
pulses.}
\end{figure}

\begin{figure}
\begin{minipage}{0.24\linewidth}
\flushright{\psfig{figure=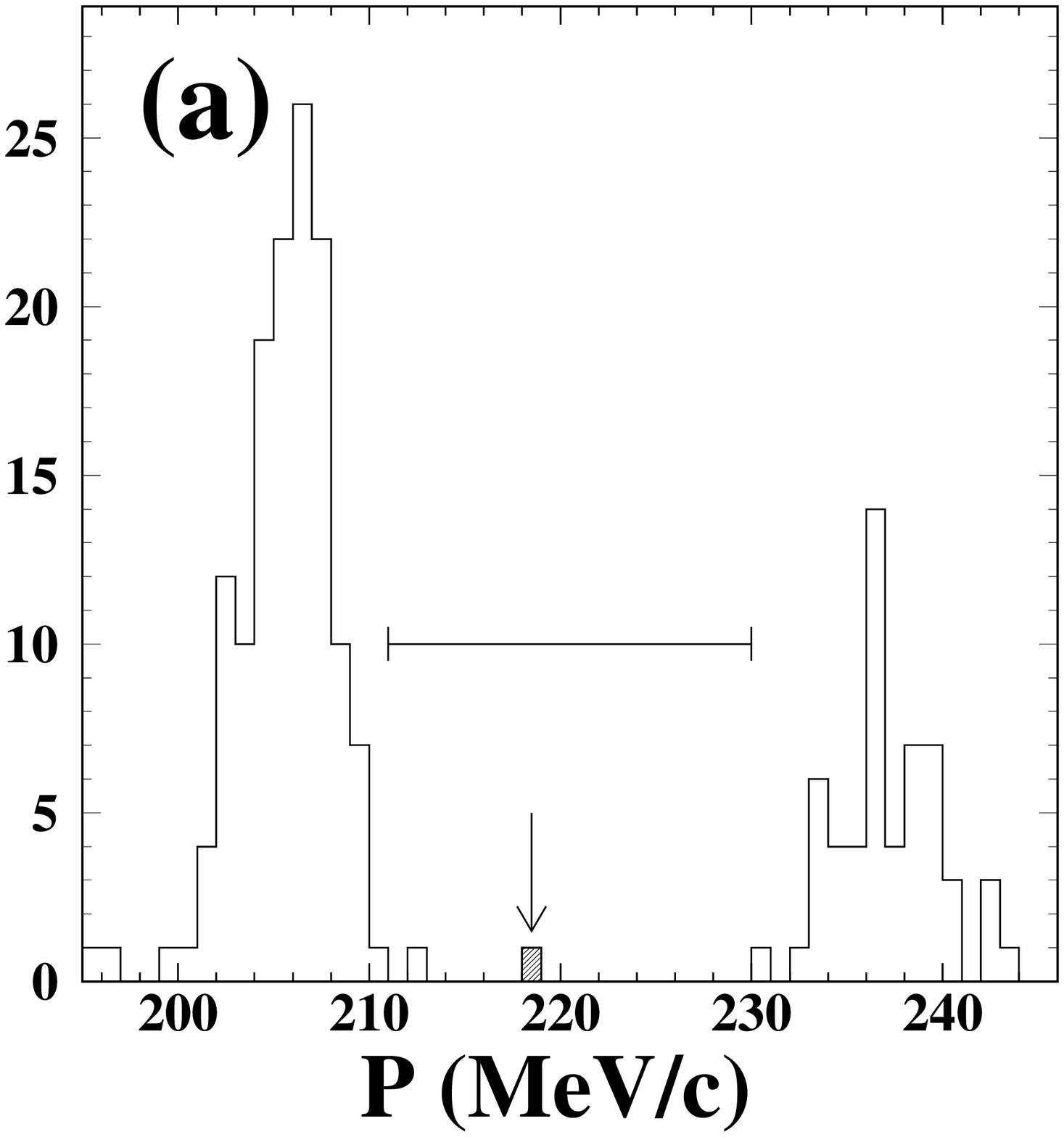,height=2.62in,width=2.62in}}
\end{minipage}\hfill
\begin{minipage}{0.24\linewidth}
\flushleft{\psfig{figure=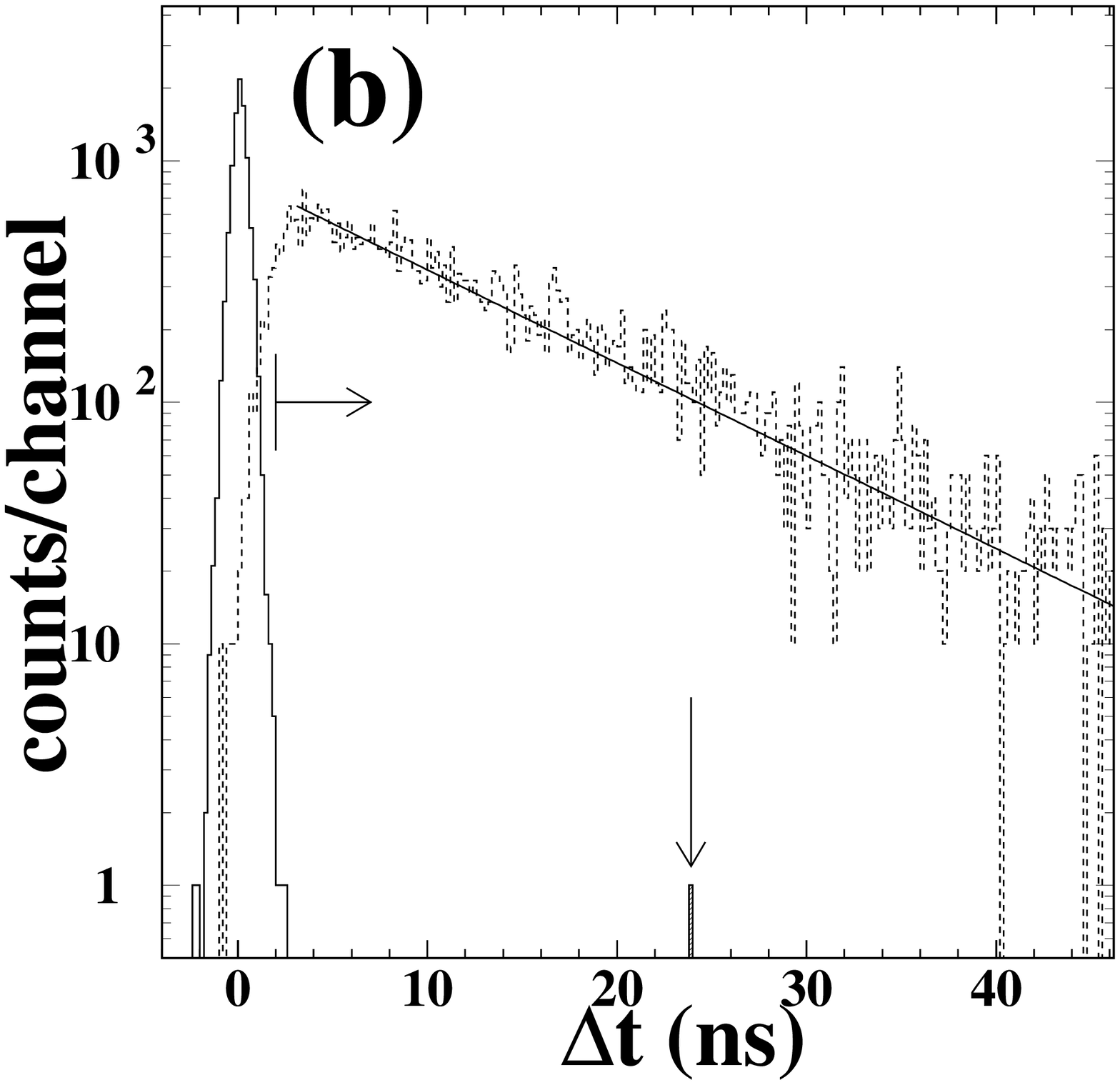,height=2.62in,width=2.62in}}
\end{minipage}\hfill
\caption{\label{evtbkg}}{Enhanced background distributions. (a) The
histogram shows the momentum spectrum with backgrounds enhanced by 
an order of magnitude by loosening the range, photon and TD particle
identification cuts. The peaks are due to $K_{\pi 2}$ and $K_{\mu 2}$.
The candidate event (vertical arrow) is shown in relation to the
accepted region (horizontal bar). (b) The time difference $\Delta t$
between the $\pi^+$ and the $K^+$ signals in the stopping target for
the candidate event (vertical arrow) and for a sample of events
identified as scattered beam pions (solid histogram). Also shown is
the delay timing cut position (horizontal arrow) and the measured time
distribution for kaon decays (dotted histogram). The straight line
shows the $K^+$ lifetime.}
\end{figure}

A reconstruction of the candidate event is shown in Fig.~\ref{event1},
and the momentum and timing of the candidate event are shown in
relation to enhanced background distributions in Fig.~\ref{evtbkg}.
Measured parameters of the event include $P=219.1\pm2.9$ MeV/$c$,
$E=118.9\pm3.9$ MeV, $R=36.3\pm 1.4$ cm, and decay times $K\to\pi$,
$\pi \to \mu$ and $\mu \to e$ of $23.9\pm0.5$ ns, $27.0\pm0.5$ ns and
$3201.1\pm 0.7$ ns, respectively.  No significant energy was observed
elsewhere in the detector in coincidence with the pion\cite{noise}.
The event also satisfied the most demanding criteria designed in
advance for candidate evaluation.  This put it in a region with an
additional background rejection factor of 10. In this region,
$b'=0.008 \pm 0.005$ events would be expected from known background
sources while 55\% of the final acceptance for $K^+ \to \pi^+ \nu
\bar\nu$~ would be retained\cite{bclean}.  Since the explanation of
the observed event as background is highly improbable, we conclude
that we have likely observed a kaon decay $K^+ \to \pi^+ \nu \bar\nu$.

To calculate the branching ratio indicated by this observation, we
used the final acceptance for $K^+ \to \pi^+ \nu \bar\nu$, $A=0.0016
\pm 0.0001^{stat} \pm 0.0002^{syst}$ derived from the factors given in
Table~\ref{acceptance}, and the total exposure of $N_{K^+}=1.49\times
10^{12}$ kaons entering the target.  Where possible, we employed
calibration data taken simultaneously with the physics data for the
acceptance calculation. We relied on Monte Carlo studies only for the
solid angle acceptance factor, the $\pi^+$ phase space factor and the
losses from $\pi^+$ nuclear interactions and decays in flight.
Figure~\ref{data2}(b) shows the simulated $R$ vs.~$E$ distribution for
$K^+ \to \pi^+ \nu \bar\nu$~ with final analysis cuts applied.  The
systematic uncertainty in the acceptance was estimated to be about
$10\%$. This was confirmed by comparing a parallel measurement of the
$K_{\pi 2}$ branching ratio to the world-average value.  If the
observed event is due to $K^+ \to \pi^+ \nu \bar\nu$, the branching
ratio is $B(K^+ \! \rightarrow \!  \pi^+ \nu \overline{\nu}) =
4.2^{+9.7}_{-3.5} \times 10^{-10}$.

The likelihood of the candidate event being due to $K^+\to \pi^+ X^0$
($M_{X^0} = 0$) is small. Based on the measured resolutions, the
$\chi^2$ CL for consistency with this hypothesis is 0.8\%.  Thus,
using the acceptance for $K^+\to \pi^+ X^0$, $A_{(K^+ \to \pi^+
X^0)}=0.0052 \pm 0.0003^{stat} \pm 0.0007^{syst}$, and no observed
events in the region $221 < P < 230$ MeV/$c$, a 90\% CL upper limit of
$B(K^+ \!  \rightarrow \!  \pi^+ X^0) < 3.0 \times 10^{-10}$ was
derived.

The observation of an event with the signature of $K^+ \to \pi^+ \nu
\bar\nu$~ is consistent with the expectations of the SM which are
centered at about $1 \times 10^{-10}$. Using the result for $B$($K^+
\to \pi^+ \nu \bar\nu$) and the relations given in ref.~\cite{BB3},
$|V_{td}|$ lies in the range $0.006<|V_{td}|< 0.06$\cite{vtd}.  E787
has recently collected additional data and the experiment is
continuing.

\acknowledgements

We gratefully acknowledge the dedicated effort of the technical
staff supporting this experiment and of the Brookhaven AGS
Department.  This research was supported in part by the
U.S. Department of Energy under Contracts No. DE-AC02-76CH00016,
W-7405-ENG-36, and grant DE-FG02-91ER40671, by the Ministry of
Education, Science, Sports and Culture of Japan and by the Natural
Sciences and Engineering Research Council and the National Research
Council of Canada.




\begin{table}
\begin{center}
\begin{tabular}{|l|l|}
~Acceptance factors& \\
\hline
~$K^+$ stop efficiency&   $0.75 $ \\
~$K^+$ decay after 2 ns&  $0.813 $ \\
~$K^+ \to \pi^+ \nu \bar\nu$~  phase space &   $ 0.162 $ \\
~Solid angle acceptance         &   $0.386 $\\
~$\pi^+$ nucl. int., decay-in-flight & $0.502$ \\
~Reconstruction efficiency &          $0.956 $ \\
~Other kinematic constraints &              $0.713 $ \\
~$\pi - \mu -e$ decay acceptance&     $ 0.247 $\\
~Beam and target  analysis &           $0.659 $\\
~Accidental loss &         $0.747 $\\
\hline
Total acceptance&   $0.0016 $\\
\end{tabular}
\end{center}
\caption{\label{acceptance}}{Acceptance factors used in the
measurement of $K^+ \to \pi^+ \nu \bar\nu$. The ``$K^+$ stop
efficiency'' is the fraction of kaons entering the target that
stopped, and ``Other kinematic constraints'' includes kinematic
particle identification and $dE/dx$ cuts.  }

\end{table}

\newpage

\newpage

\newpage


\begin{references}
\bibitem{bkgold} An order of magnitude improvement in background
suppression was obtained relative to ref.\cite{pnn96}. The $K_{\mu 2}$
($K_{\mu 2 \gamma}$) and $K_{\pi 2}$ backgrounds were reduced due to a
factor of two better momentum resolution (see \cite{utc}) and to
improved photon detectors (see \cite{csi}). The beam and CEX
backgrounds were reduced, in part, due to improved timing resolutions
(see \cite{ccd}).
\bibitem{noise} There are a few scattered hits off the track in the
drift chamber which are consistent with being due to random noise.
\bibitem{bclean} The background estimates in the region with the tightest
cuts were $b_{{K_{\mu 2}}} = 0.004 \pm 0.004$, $b_{{K_{\pi 2}}} =
0.001 \pm 0.001$, $b_{Beam} = 0.002\pm 0.002$ and $b_{CEX} = 0.002 \pm
0.002$.
\bibitem{vtd} These limits include the values of $|V_{td}|$ consistent
with the unitarity of the quark mixing matrix, the range of $B(K^+ \!
\rightarrow \! \pi^+ \nu \overline{\nu})$ given in the text, and the
ranges of $V_{cb}$, $m_c$, $m_t$, and $\Lambda_{\overline{MS}}$ given
in ref~\cite{bf}.  However, they are independent of (and consistent
with) information on $|V_{ub}/V_{cb}|$, $B-{\overline B}$ mixing, or
CP-violation in the $K$ system used in other estimates of the range of
$|V_{td}|$.
\end{references}
\end{document}